\documentclass[headsepline,a4paper,twoside,11pt,smallheadings]{scrartcl}

\typearea[16mm]{13}

\usepackage{times}
\usepackage{mathptm}

\usepackage{graphics}
\usepackage{natbib}
\bibpunct[,]{(}{)}{;}{a}{}{,}
\usepackage[small,bf]{caption}

  \renewenvironment{thebibliography}[1]{%
    \begin{oldthebibliography}{#1}%
      \setlength{\parskip}{0ex}%
      \setlength{\itemsep}{0ex}%
  }%
  {%
    \end{oldthebibliography}%
  }

\title{Halo and Galaxy Formation Histories from the Millennium Simulation:\\
  {\em \LARGE Public release of a VO-oriented and SQL-queryable database for
  studying the evolution of galaxies in the $\Lambda$CDM cosmogony}}

\author{Gerard Lemson\footnote{German Astrophysical Virtual
Observatory, Max Planck Institutes 
for Astrophysics and for
extraterrestrial Physics}
\footnote{{\it email:} gerard.lemson@mpe.mpg.de}
~ and the Virgo Consortium}

\date{{\bf 31 July 2006}\vspace*{-1.1cm}}

\begin{document}
\maketitle

\abstract{{\bf ABSTRACT} The Millennium Run is the largest simulation of the
formation of structure within the $\Lambda$CDM cosmogony so far carried
out. It uses $10^{10}$ particles to follow the dark matter distribution in a
cubic region 500$h^{-1}$Mpc on a side, and has a spatial resolution of 5
$h^{-1}$kpc. Application of simplified modelling techniques to the stored
output of this calculation allows the formation and evolution of the $\sim
10^7$ galaxies more luminous than the Small Magellanic Cloud to be simulated
for a variety of assumptions about the detailed physics involved.  As part of
the activities of the German Astrophysical Virtual Observatory we have used a relational database
to store the detailed assembly histories both of all the
haloes and subhaloes resolved by the simulation, and of all the galaxies that
form within these structures for two independent models of the galaxy
formation physics. 
We have created web applications that allow users to query these databases remotely
using the standard Structured Query Language (SQL).
This allows easy access to all properties of the
galaxies and halos, as well as to the spatial and temporal relations between
them and their environment. Information is output in table format compatible with standard Virtual
Observatory tools and protocols. With this announcement we are making these structures fully
accessible to all users. Interested scientists can learn SQL, gain familiarity with the 
database design and test queries
on a small, openly accessible version of the Millennium Run (with volume 1/512
that of the full simulation). They can then request accounts to run similar
queries on the databases for the full simulations. }
\newpage

\section{The Millennium Run}
The last few years have seen the establishment of a standard model for the
origin and growth of structure in the Universe, the so-called $\Lambda$CDM
cosmogony. In this model small density fluctuations are generated during an
early period of cosmic inflation and first become directly observable on the
last scattering surface of the Cosmic Microwave Background when the Universe
was about 400,000 years old.  Since this time the fluctuations have grown
steadily through the gravitational effects of a dominant Dark Matter component
composed of some weakly interacting particle yet to be detected directly on
Earth. As fluctuations become nonlinear, larger and larger objects collapse,
giving rise to the galaxies and galaxy clusters we see today. This process has
recently been modified as Dark Energy has come to dominate the cosmic energy
density, accelerating the cosmic expansion and reducing the rate of structure
growth. A major effort is currently underway, testing this paradigm and
measuring its parameters. A parallel effort explores galaxy and cluster
formation in this model in order to understand the physical processes which
shaped observed systems.

The Millennium Run, completed in summer 2004 at the Max Planck Society's
supercomputer centre in Garching, is part of the programme of the Virgo
Consortium\footnote{http://www.virgo.dur.ac.uk} and is intended as a tool to
facilitate this second effort. It uses $10^{10}$ particles of mass $8.6\times
10^8h^{-1}$M$_\odot$ to follow the evolution of the dark matter distribution
within a cubic region of side $500h^{-1}$Mpc from $z=127$ until $z=0$. The
cosmological parameters assumed are $\Omega_m=\Omega_{dm}+\Omega_b=0.25$,
$\Omega_b=0.045$, $\Omega_\Lambda=0.75$, $h=0.73$, $\sigma_8=0.9$ and $n=1$
with standard definitions for all quantities. The initial density fluctuations
correctly account for the oscillatory features introduced by the baryons, but
the simulation follows the dark matter only, supplementing the mass of the
simulation particles to account approximately for the neglected baryons.

The simulation was carried out using a modified version of the publicly
available code GADGET-2 \citep{Springel2005b}. The positions and 
velocities of all simulation particles were stored at 63 times spaced 
approximately logarithmically from $z=20$ to the present day. For each
of these dumps the algorithm SUBFIND \citep{Springel2001b} was used
to identify all self-bound halos containing at least 20 particles and all
self-bound subhalos within these halos down to the same mass limit.
Merger trees were then built linking each halo and its substructures
at the final time to the objects at earlier times from which they formed.
These trees are the input to the final stage of post-processing. This
simulates the formation of the galaxies in all or a part of the volume
by following simplified treatments of the baryonic physics within each tree,
starting at early times and integrating down to $z=0$. More detailed
descriptions of the simulation itself and of this post-processing can be
found in \citet{Springel2005a}.

Several different galaxy formation models have already been implemented on
this structure by the Garching and Durham groups. The model used in
\citet{Springel2005a} to present some initial clustering and evolution results
is essentially identical to that described and explored in considerably more
detail by \citet{Croton2006a}. The model used by \citet{DeLucia2006a} to study
elliptical galaxy evolution is similar in most aspects, but differs in its
treatment of feedback from star formation.  In their study of brightest
cluster galaxies \citet{DeLucia2006b} use a model with a different assumed IMF
for star formation, an improved scheme for tracking halo central galaxies, but
the same feedback scheme as \citet{Croton2006a}. The model independently
developed in Durham and presented in \citet{Bower2006} differs from these
Garching models in many ways. The scheme for building merger trees from the
halo/subhalo data is different in detail, as are many of the modelling
assumptions made to deal with the baryonic physics, most notably, perhaps,
those associated with the growth of and the feedback from supermassive black
holes in galaxy nuclei. In this public release we are initially making
available the galaxy populations produced by the models of the
\citet{DeLucia2006b} and \citet{Bower2006} papers.

The data on the halo/subhalo and galaxy populations which have been produced by
this effort can be used to address a very wide range of questions about galaxy
and structure evolution in the now standard model. In the 13 months since {\it
Nature} published the first Millennium Run paper \citep{Springel2005a} a
further 24 papers have appeared on the preprint server using data derived from
the simulation. Some of these are concerned with issues of dark matter
structure \citep{Gao2005a,Harker2006a,Gao2006a}. Others build and test galaxy
formation models, exploring the requirements for reproducing various aspects
of the observed properties of galaxies and AGN \citep{Croton2006a,Croton2006b,
Bower2006,DeLucia2006a,Croton2006c,Wang2006a,DeLucia2006b}. Yet others
concentrate on aspects of large-scale structure and galaxy clustering
\citep{Croton2006d,Noh2006a,Lee2006a} and on cluster structure and
gravitational lensing \citep{Hayashi2006a,Moeller2006a,Weinmann2006a,
Natarajan2006a}. A number of authors have used galaxy catalogues from the
Millennium Run as a point of comparison in primarily observational papers
\citep{Kauffmann2006a,Patiri2006a,Einasto2006a,Rudnick2006a,Bernardi2006a,
Conroy2006a,Li2006a}. Finally, the data have been used to illustrate the
current state-of-the-art in a review of the field of large-scale structure
\citep{Springel2006a}. The goal of this release is to facilitate further such
use of Millennium Run data products by making them conveniently and publicly
available over the Web.

\section{The Databases}

The stored raw data from the Millennium Run, the positions and velocities of
all $10^{10}$ particles in the initial conditions and at each of the 63 later
output times, have a total volume of almost 20TB. This is so large that
general public access and/or manipulation over the internet is not currently a
viable possibility.  As a result, projects which require access to the full
particle data (e.g. ray-tracing projects for gravitational lensing
applications) are only practicable in collaboration with Virgo scientists in
Garching or Durham, where copies of the full data are stored. The Virgo
Consortium welcomes suggestions for such joint projects and will try to
accommodate them as far as they overlap with the interests of scientists at
one of the Virgo institutions and do not conflict with existing projects.

Many projects, however, including the great majority of those listed at the
end of \S 1, can be carried out using products from our Millennium Run
post-processing pipeline. Only $O$(130GB) are needed to store the information
provided by our halo-subhalo analysis including the tree structure which
describes the assembly history of all objects. A database with the
corresponding galaxy information from one of our galaxy formation simulations
is roughly twice as large because of the larger number of objects and the
larger number of attributes ascribed to each. The variety of these attributes
and the complexity of the relations between them motivate the use of
relational databases, whose query engines allow complex questions to be
phrased in the standard Structured Query Language (SQL) and executed in
optimal fashion.  These tools have become standard in the Virtual Observatory
community as a means for promoting efficient and user-friendly data-mining
within large observational databases.  A major task for the German
Astrophysical Virtual Observatory (GAVO) has been to adapt these tools for
large theoretical (simulation) databases, where issues of format and quality
control are less difficult than for observational archives, but where
relationships can be considerably more complex, primarily because of the
addition of the time dimension.

The relational database structure and the online query interface which we have
set up for Millennium Run data products are modelled closely on the relevant
parts of the very successful SkyServer system set up for the Sloan Digital Sky
Survey.\footnote{http://cas.sdss.org/dr5/en/}  Within
such a database information is stored in the form of tables where rows
correspond to individual objects and columns to attributes of those objects
(e.g. position, velocity, mass, angular momentum, size, flattening, type ,
luminosity, colour, indices specifying relations to other objects...).  The
web interface allows users to formulate their scientific questions as SQL
queries operating on these tables in a relatively simple way and to submit
them remotely over the internet for execution on the database server (located
at present in Garching, with a mirror to be set up in the near future in
Durham). The subsequent search of the database is optimised as far as possible
for ``typical'' queries. Results are returned to the user over the internet as
tables in one of a number of formats which can then be fed into standard VO or
other graphics packages or can be further manipulated by users with their own
software.

The entry point for our public release of Millennium products is\\
http://www.mpa-garching.mpg.de/Millennium\\ This top page gives a brief
introduction to the Millennium Run as well as links to images and movies, to
papers which have used the data, and to the pages on the GAVO site which
describe in detail the database structure, the SQL and the procedures for
accessing and downloading data. Data on the so-called ``milli-Millennium'' run
(hereafter milli-M) can be accessed from these pages immediately.  This is a
simulation which is identical to the main run in all aspects except that it is
carried out in a cubic region of side $62.5h^{-1}$Mpc. It thus has 1/512 of
the physical volume and its databases are 512 times smaller than those of the
Millennium Run itself. Any query can be executed on the milli-M databases
directly from this page, provided it executes on the host computer in less
than 30 seconds. The first 10,000 lines of any output are returned to the
user. These restrictions are intended to avoid inadvertently tying up the host
or the internet link while developing SQL expertise.

Once users can execute their queries efficiently on the milli-M databases,
they can apply for password-protected accounts as specified on the web-page in
order to carry out the corresponding queries on the main database. This
two-stage system is intended to allow monitoring of usage patterns and to
prevent accidental abuse by inexperienced users.

At present databases are accessible for the (sub)halo data and for
galaxy data from the models of \citet{DeLucia2006b} and \citet{Bower2006}.
identical models and attribute lists are used for the milli-M and full
Millennium versions of these databases.  Further galaxy models and further
data products may be added as they are generated. Examples of scientific
queries that can be formulated simply and executed efficiently with the
present SQL engine include:
\begin{itemize}
\item Find all halos (or galaxies) in a given part of the simulation at a 
      given time and in a given mass range
\item Find all companion halos (or galaxies) in some given range of
      separations from this previous set of objects
\item Find the number of galaxies at a given time in each of a series of narrow
      luminosity bins (e.g. the luminosity function)
\item Find the number of galaxies in high mass halos at a given time in such
      luminosity bins (e.g. the cluster luminosity function)
\item Find all resolved progenitor halos at redshift 3 of high-mass $z=0$ halos
\item Find all galaxies at redshift 3 which are progenitors of the central 
      galaxies of high-mass $z=0$ halos
\item Find the halo masses of all $10^{11}$M$_\odot$ galaxies at $z=3$ and
      determine the fraction which are central galaxies of these halos
\item Find the $z=0$ descendents of all redshift 3 galaxies with stellar
      mass above $10^{11}$M$_\odot$ or with star formation rate above
      10M$_\odot$/yr
\item Find all halos (or galaxies) which have undergone a major merger since
      the previous stored output time
\end{itemize}
Clearly this capability allows a very broad range of scientific issues to be
addressed in a straightforward way.  Some of these are currently being studied
by scientists associated with the Virgo Consortium, but we hope that this
release will encourage others to use the exceptional statistics provided by
the Millennium Run to explore how galactic and dark matter structures evolve
in the current $\Lambda$CDM paradigm. In particular, closer comparison with a
wide range of observational data should indicate how our present simple models
for the formation of galaxies and AGN need to be modified to correspond better
with reality.

\section*{Acknowledgements} The German Astrophysical Virtual Observatory
(GAVO) is funded by the German Federal Ministry for Education and Research.
We are grateful to the Rechenzentrum Garching, the supercomputer centre
of the Max Planck Society, for their help in executing the Millennium Run.
We are very grateful to Alex
Szalay and his team at Johns Hopkins University for discussions and
help in the design and implementation of the database.
Simon White is grateful to the Kavli Institute of Theoretical Physics (supported
in part by the National Science Foundation under grant PHY99-07949) for
their hospitality during the writing of this announcement.

\bibliographystyle{phd} 
\bibliography{proposal}

\end{document}